\documentclass[letterpaper,twocolumn,aps,showpacs,12point,prl]{revtex4-1}
\usepackage{amsmath}
\usepackage{amssymb}
\usepackage{graphicx}

\makeatletter

\@ifundefined{textcolor}{}
{
 \definecolor{BLACK}{gray}{0}
 \definecolor{WHITE}{gray}{1}
 \definecolor{RED}{rgb}{1,0,0}
 \definecolor{GREEN}{rgb}{0,1,0}
 \definecolor{BLUE}{rgb}{0,0,1}
 \definecolor{CYAN}{cmyk}{1,0,0,0}
 \definecolor{MAGENTA}{cmyk}{0,1,0,0}
 \definecolor{YELLOW}{cmyk}{0,0,1,0}
}

\usepackage{color}

\usepackage{graphics}

\newcommand{\ket}[1]{|#1\rangle}
\newcommand{\M}[4]{\begin{pmatrix} #1 & #2 \cr #3 & #4 \end{pmatrix}}

\makeatother

\begin{document}

\title{Randomized benchmarking of single qubit gates in a 2D array of neutral atom qubits}

\author{T. Xia, M. Lichtman, K. Maller, A. W. Carr, M. J. Piotrowicz, L. Isenhower, and M. Saffman}

\affiliation{Department of Physics, University of Wisconsin, 1150 University Avenue,
Madison, WI 53706}
\begin{abstract}
We characterize single qubit Clifford gate operations
with randomized benchmarking in a 2D array of neutral atom qubits, and demonstrate  global and site selected gates with high fidelity. An average  fidelity of $F^2=0.9983(14)$ is measured for global microwave driven gates applied to a 49 qubit array.
 Single site gates are implemented with a  focused laser beam to Stark shift the microwaves into resonance at a selected site. 
At Stark selected single sites we observe $F^2=0.9923(7)$ and an average spin flip crosstalk error at other sites of $0.002(9)$. 

\end{abstract}

\pacs{03.67.-a, 31.10.Jk, 42.50.Dv, 03,67.Lx.}

\maketitle

Qubits encoded in hyperfine states of neutral atoms are one of several 
promising approaches for scalable  implementations of quantum information processing\cite{Ladd2010}. In this letter we demonstrate and characterize single qubit gate operations in a 2D array of up to 49 neutral atom qubits encoded in long lived hyperfine states. Using a microwave field we implement arbitrary Bloch sphere rotations on either the entire 2D qubit array in parallel, or on single sites that are selected by an auxiliary Stark shifting laser beam\cite{CZhang2006}.
Single atom qubits are stochastically loaded into the array, with an average of 29 sites filled for the data reported here. This is the largest number of  individually controllable 
qubits for which quantum gate operations have been characterized to date. The fidelity of the global operations, site selected operations, and crosstalk during site selected operations are quantified using randomized benchmarking (RB)\cite{Knill2008}.  

Control of individual qubits in a spatially extended array is an important capability in ongoing efforts to develop scalable quantum processors. Atomic qubits encoded in hyperfine ground states can be controlled with optical or microwave frequency fields. Optical fields can be tightly focused to address individual qubits as has been demonstrated in several experiments\cite{Nagerl1999,*Yavuz2006,*Knoernschild2010,*Labuhn2014,*Merrill2014,*Crain2014}. When the spatial separation of qubits is comparable to, or less than,  the optical wavelength, addressing by focusing alone is not sufficient to suppress crosstalk to neighboring sites. Addressing with subwavelength resolution can be achieved using quantum interference techniques\cite{Miles2013} or by using an additional external field gradient to select a desired site. This latter method has been implemented with magnetic field gradients\cite{Schrader2004,*Johanning2009}, or with auxiliary Stark shifting optical beams in conjunction with microwave fields\cite{CZhang2006,Weitenberg2011,Lundblad2009,Lee2013}. The use of microwave fields for qubit control is particularly convenient since both the global rotations which are the starting point for many quantum algorithms, as well as single qubit control needed for gates, can be implemented with the same control hardware. 

In this letter we demonstrate global and site selected single qubit gates using microwave drive fields  with a tightly focused Stark shift beam.  We derive optimal values for the Stark shift which minimize crosstalk to other sites. Arbitrary rotations on the Bloch sphere are implemented using variable length and phase microwave pulses. In contrast to previous experiments which used either adiabatic pulses that do not provide full control on the Bloch sphere\cite{Weitenberg2011}, or spatially periodic Stark shifting techniques which do not address single sites\cite{Lundblad2009,Lee2013}, we demonstrate full control at single sites of a 2D qubit array.
Using RB techniques we characterize the fidelity of  Clifford group gates,  as well as the crosstalk during site selected gates\cite{Gambetta2012,*Kelly2014}.

RB was introduced in \cite{Knill2008} as an efficient approach for characterization of quantum gate fidelities. It has several advantages compared to full tomography including resource requirements that scale linearly with the number of qubits and the capability of distinguishing gate errors from state preparation and measurement (SPAM) errors. RB has been used for characterization of one- and two-qubit gates, as well as quantum processes,  on a variety of qubit platforms including 
ions\cite{Knill2008,Biercuk2009,*Brown2011,*Gaebler2012,*Gaebler2012err},
nuclear magnetic resonance\cite{Ryan2009},
 superconductors \cite{Chow2009,*Magesan2012,*Barends2014b},
 neutral atoms \cite{Olmschenk2010,Smith2013,Lee2013},
and quantum dots\cite{Veldhorst2014}. We encode qubits in the 
Cs clock states with  $|0\rangle\equiv|f=3,m_f=0\rangle$, $|1\rangle\equiv|f=4,m_f=0\rangle$. Our implementation of RB uses the 
complete set of 24 Clifford gates $\mathcal C_1$ . These are generated from the set $\{I,R_j(\pm\pi/2),R_j(\pi)\}$ where $R_j(\theta)=e^{-\imath\theta\sigma_j/2}$ with $\sigma_j$ Pauli matrices about axes $j=x,y,z$. We use constant amplitude pulses of microwave radiation resonant with the $\omega_q=2\pi\times 9.19 ~\rm GHz$ $\ket{0}\leftrightarrow\ket{1}$ clock transition for $R_x$ rotations. Phase shifting the microwaves provides $R_y$ rotations. $R_z$ operations are implemented by composing $x$ and $y$ axis rotations. The microwave pulses used for each Clifford gate are listed in the supplementary material.

\begin{figure}[!t]
\includegraphics[width=0.99\columnwidth]{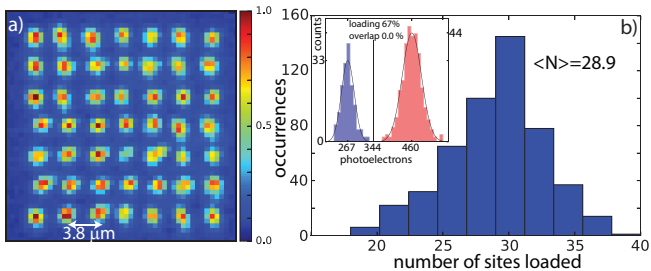}
\vspace{-.3cm}
\caption{(color online) Qubit array measurements. a) False-color fluorescence image of single atom qubits filtered with an independent component analysis. Each pixel views a region $0.62\times 0.62~\mu\rm m$  The image is a composite averaged over 500 exposures.   b) Histogram of the number of occupied sites for 500 array loading measurements.  The inset shows the single atom photoelectron counts at a single site for 2000 measurements. The average overlap of the Gaussian fits at all 49 sites was 0.0004.}
\label{fig.array}
\end{figure}

The main elements of the atomic experiment are as described in \cite{Piotrowicz2013}. In brief, a two-dimensional array of blue detuned optical traps is defined using 780 nm light projected into a pyrex ultrahigh vacuum cell. The $7\times7=49$ site array has a  $3.8~\mu\rm m$ site to site spacing and trap depths of $\sim 400 ~\mu\rm K$ for Cs atoms.  The Cs atoms are collected in a 2D cooling  region, transferred to the pyrex cell with a push beam,  and then trapped and cooled in a 3D magneto-optical trap (MOT). The array is then turned on, the MOT quadrupole field is turned off,  and the captured atoms are cooled to 5-10 $\mu\rm K$ using polarization gradient cooling. About 5  ms of near resonant light at 852 nm is used to invoke photo-assisted collisions which remove atoms in multiply occupied  sites. After this preparation step multiple atoms are not observed at any site.   
The presence of an atom is detected by fluorescence imaging as shown in Fig.\ref{fig.array}.
We observe single atom loading rates approaching 70\% at a few sites, which is suggestive of repulsive light assisted collisions\cite{Carpentier2013}. On average we load 60\% of the array sites with a single atom in each experimental run.  

An initial fluorescence image reveals which sites are loaded with qubits for each experimental run. 
The atoms are then optically pumped into $\ket{1}$ using $\pi$ polarized 894 nm light resonant with $\ket{6s_{1/2},f=4} \rightarrow \ket{6p_{1/2},f=4}$ and $\pi$ polarized 852 nm repump light resonant with $\ket{6s_{1/2},f=3} \rightarrow \ket{6p_{3/2},f=4}$. The quantization axis is perpendicular to the plane of the array and is defined by a 0.15 mT magnetic bias field.  After the quantum gate operations described below a state sensitive measurement is performed. To measure the probability of $\ket{0}$ we push out atoms in $f=4$ with unbalanced resonant light pressure ($\ket{6s_{1/2},f=4} \rightarrow \ket{6p_{3/2},f=5}$) and then measure the presence of an atom by integrating the fluorescence from  
MOT light with detuning $-7\gamma_{6p_{3/2}}$ and resonant saturation parameter 3.3 for 20 ms.  
This results in high fidelity discrimination of the qubit states as is seen in Fig. \ref{fig.array}b).   To measure the probability of $\ket{1}$ we apply a $R_x(\pi)$ microwave pulse before the push out and fluorescence measurement. All sites in the array are measured in parallel using camera detection of the fluorescence. 
The dominant error in state measurement is the
small probability of  transferring an atom from $f=4$ to $f=3$ during the pushing out step.

\begin{figure}
\includegraphics[width=0.99\columnwidth]{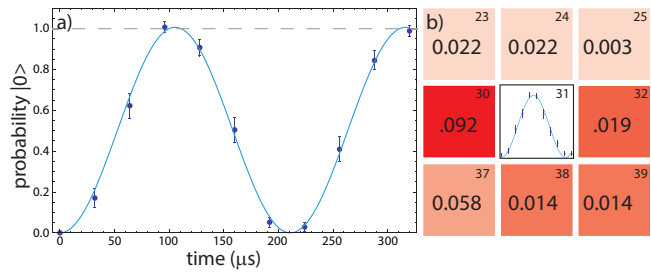}
\vspace{-.3cm}
\caption{(color online) a) Microwave Rabi oscillations at a single site during global addressing. Each point is the average of 50  measurements and the fitted Rabi frequency is $\Omega=2\pi\times4.74$ kHz. b) Oscillations on site 31 using Stark addressing with the average probability of measuring a neighboring site in the wrong state due to crosstalk during a 3$\pi$ oscillation indicated numerically. The Rabi frequency for this data was $\Omega=2\pi\times8.5$ kHz and the sites are numbered from 0 in the upper left corner to 48 in the lower right corner. }
\label{fig:rabi}
\end{figure}

In preparation for quantum gate experiments the qubit array characteristics are measured. Array averaged values are 17 s for the $1/e$ atom lifetime, 0.59 s for $T_1$ and 14 ms for $T_2^*$. The $T_2^*$ value, which is measured using microwave Ramsey spectroscopy, is dominated by magnetic noise and finite temperature motional effects\cite{Kuhr2005}. The 9.19 GHz microwave source is
locked to a GPS disciplined crystal oscillator. The frequency is slightly shifted from the free space Cs clock transition due to the magnetic bias field and the $\sim 500~\rm  Hz$ light shift at the center of each trapping site. 
A maximum $T_2^*$ of $\sim 50 ~\rm ms$ has been observed at a few sites which we attribute to variations in the cooling efficiency and  atom temperature. We anticipate that the $T_2^*$ value can be substantially improved in future experiments using trap compensation techniques\cite{Carr2014}.

We proceed with RB experiments to measure the
fidelity of  single qubit gate operations with microwave radiation from a horn external to the vacuum cell driving all qubits in parallel.
Random Clifford gate sequences of length $\ell$ are generated with  each gate  chosen uniformly from $\mathcal C_1$. The average pulse area per gate was $7\pi/4$ (see supplemental material). We start with all qubits in $\ket{1}$. 
At the end of each sequence we add a final gate which, in the absence of errors, should transfer the qubits to $\ket{0}$. In the presence of depolarization errors the probability of measuring $\ket{0}$ is 
\begin{equation}
P_{\ket{0}}=\frac{1}{2}+\frac{1}{2}(1-d_{\rm if})(1-d)^{\ell}.\label{eq:fid}
\end{equation}
Here $d_{\rm if}$ is the depolarization probability associated with state preparation, the final transfer gate, and state measurement, while $d$ is the average depolarization of a Clifford gate. Using the standard definition\cite{Nielsen2000} of the fidelity of two density matrices $\rho,\rho'$ given by $F(\rho,\rho')={\rm Tr}[\sqrt{\sqrt\rho\rho'\sqrt\rho}]$ one can readily show that the square of the average gate fidelity is $F^2=1-d/2.$ The quantity $F^2$ is equal to the average fidelity of a Clifford gate $F_a$ introduced in \cite{Knill2008}.

\begin{figure}[!t]
\includegraphics[width=.9\columnwidth]{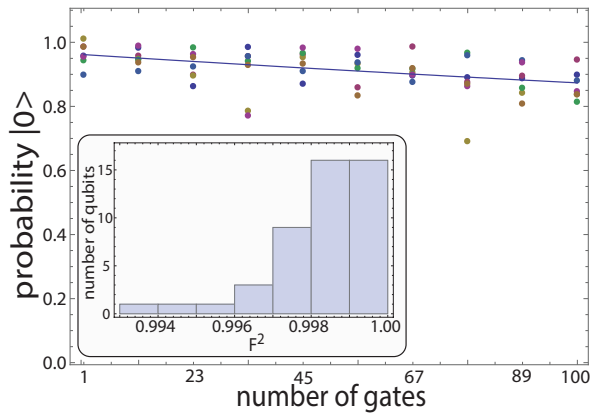}
\vspace{-.4cm}
\caption{(color online) Probability of measuring the correct output state at site 27  of the array for 7 RB sequences. Each sequence was truncated at 10
different lengths $\ell=\{1, 12, 23, 34, 45, 56, 67, 78, 89, 100\}$. Each data point is an average of 50 measurements.  The inset shows a histogram of  gate fidelities for 47 of the 49 array sites. Two sites were dropped due to poor loading statistics. }
\label{fig:global_ROI27}
\end{figure}

We applied 7 randomized Clifford gate sequences for all the trapping
sites. Representative data from a single site are shown in Fig. \ref{fig:global_ROI27} together with a histogram of $F^2$ across the array obtained by extracting $d$ from fits to Eq. (\ref{eq:fid}) at each site.   The results are summarized in Table \ref{tab:results}. The highest fidelity seen at any site was 0.9999(3) with an array average of 0.9983(14). These values are comparable to the highest fidelity neutral atom gates reported to date\cite{Olmschenk2010}, who reported a global average RB fidelity of 0.99986(1). An indication of where these experimental results stand in relation to theoretical thresholds for   fault tolerant quantum computing can be found by consulting Table 8 in Ref. \cite{Devitt2013}.

In order to understand the sources of the observed errors we simulated the RB experiment 
allowing for detuning from $\omega_q$ by up to 100 Hz, which corresponds to the average differential
 Stark shift of the trapped qubit states across the array. We also included  pulse length timing errors of up to 200 ns corresponding to 0.2\%  drifts of the microwave power. Accounting for these imperfections predicts gate errors several times smaller than those observed.  Including a density matrix coherence  decay
factor\cite{Kuhr2005} $\alpha(t,T_{2}^{\ast})=1/2+(1/2)/[1+0.95(\frac{t}{T_{2}^{\ast}})^{2}]^{3/2}$ we estimate 
$\langle F^2\rangle=1-\langle d\rangle/2 = 1-[1-\alpha (\langle t \rangle_{\mathcal C_1}, T_2^\ast)]/2$. Putting $\langle t \rangle_{\mathcal C_1}=\langle \theta\rangle_{\mathcal C_1}/\Omega=(7\pi/4)/(2\pi\times 4.74~\rm kHz)=185~\mu\rm s$ and $T_2^*=2.7~\rm ms$ 
we recover the observed $\left\langle F^2 \right\rangle_{\rm 47~sites} = 0.9983$ from Table \ref{tab:results}.  The median $T_2^*$ observed in the array is over twice longer at 7.0 ms. We conclude that the factors limiting the gate 
fidelity found from RB experiments are a combination of finite $T_2^\ast$ which could be improved using echo techniques or trap compensation\cite{Carr2014}, and errors in the pulse length and detuning.

\begin{table}[!t]
\caption{Results of RB fidelity measurements for global (first five rows) and single site addressing. The last three rows are
$\langle E_{\rm xt}\rangle$ the average crosstalk error on the entire array, $\langle E_{\rm xt}\rangle_{\rm  nn}$ the average crosstalk for the nearest neighbor sites, and $\langle E_{\rm xt}\rangle_{\rm   -nn}$ the array averaged crosstalk excluding the nearest neighbor sites. }
\vspace{-.4cm}
\begin{center}
\begin{ruledtabular}
\begin{tabular}{l|c}
$\left\langle d_{\rm if}\right\rangle_{47~\rm sites}$ & $0.092\pm 0.066$ \\[1 pt]
\hline
$\left\langle d\right\rangle_{47~\rm sites}$ &$0.0035\pm 0.0027$ \\[1 pt]
\hline
$\left\langle F^2\right\rangle_{47~\rm sites}$ &$0.9983\pm 0.0014$ \\[2 pt]
\hline
$F^2_{\rm min}$ & $0.9939\pm 0.0007$ \\
\hline
$F^2_{\rm max}$ &  $0.9999\pm 0.0003$ \\
\hline
\hline
$F^2_{\rm single~ site}$ &  $0.9923\pm 0.0007$  \\[1 pt]
\hline$\langle d_{\rm if}\rangle_{\rm xt}$ & $0.037\pm 0.027$  \\
\hline
$\langle E_{\rm xt}\rangle$ & $0.002\pm 0.009$  \\
\hline
$\langle E_{\rm xt}\rangle_{\rm nn}$ & $0.014\pm 0.02$  \\
\hline
$\langle E_{\rm xt}\rangle_{\rm  -nn}$ & $0.0005\pm 0.001$  \\
\end{tabular}
\end{ruledtabular}
\end{center}
\label{tab:results}
\vspace{-.8cm}
\end{table}

To perform gates on individual qubit sites we detuned the microwave frequency $\omega$ by $\delta=\omega-\omega_q\simeq 2\pi\times 33 ~\rm kHz$. This detuning  suppresses  the microwave qubit rotation by a factor scaling as $\Omega^2/\delta^2.$ We then selected a desired site using a tightly focused 459 nm beam with $1/e^2$ intensity radii of $w_x=3.2$, $w_y=2.7~\mu\rm m$  detuned by $\Delta_{\rm S}=2\pi\times 20~\rm GHz$ from the $\ket{6s_{1/2},f=4}\leftrightarrow \ket{7p_{1/2},f=4}$ transition. The beam size was chosen as a compromise between  tight focusing which gives small crosstalk to neighboring sites, and loose  focusing which reduces sensitivity to beam misalignment on the target site. The intensity of the 459 nm beam was adjusted such that the induced differential Stark shift of states $\ket{0},\ket{1}$ was set equal to $\delta$ to bring a selected site into resonance\cite{CZhang2006}. The 459 nm Stark beam was $\sigma_+$ polarized and propagated normal to the plane of the array. Using a pair of orthogonal acousto-optic modulators the Stark beam could be scanned to a desired qubit site with a switching speed under $0.5~\mu\rm s$. 

The choice of detuning $\delta$ for single site gates is a trade off between less than perfect suppression of the microwave field at small 
$\delta$ and excessive photon scattering from the Stark beam at large $\delta$. For a given value of $\delta$ the photon scattering can be reduced by working at large optical detuning $\Delta_S$, but not completely eliminated since for large $\Delta_S$,  $\delta\sim \omega_q/\Delta_S^2$ and tends to zero. An optimized working point which reduces the need for large $\delta$ can be found by choosing a detuning for which the off-resonant coupling to non-selected sites gives a pulse area which is a multiple of $4\pi$ and therefore does not disturb the qubit states. For a pulse  area of $\theta_R$ on the targeted qubit the condition for minimal disturbance of non-targeted sites is $\delta/\Omega=\left(n^2 16 \pi^2/\theta_R^2-1 \right)^{1/2}$ with $n$ an integer. Thus the leakage error should have a first local minimum for a $\pi$ pulse  at $\delta/\Omega\simeq \sqrt{15}.$ 

\begin{figure}[!t]
\begin{centering}
\includegraphics[width=.9\columnwidth]{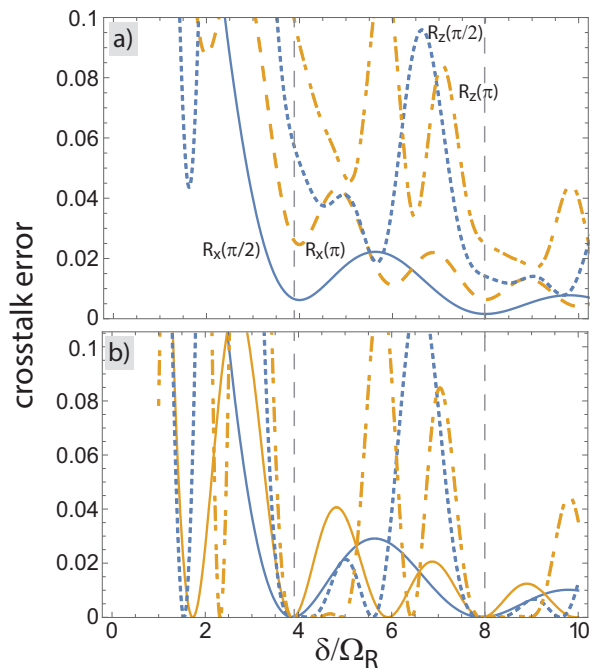}
\par\end{centering}
\vspace{-.4cm}
\caption{(color online) Crosstalk error  for $R_x(\pi/2)$(blue solid line), $R_x(\pi)$(yellow dashed line),
$R_z(\pi/2)$(blue dotted line), and $R_z(\pi)$(yellow dashed-dotted line) rotations. The vertical dashed lines mark values of $\sqrt{16n^2-1}$. Panel a) is the error from Eq. (2) averaged over input states and panel b) is the error for the initial state $\ket{1}$.   } 
\label{fig:state_fidelity}
\end{figure}

This estimate can be verified by a calculation which averages over all possible states of the non-targeted qubits. Let the initial state be  
 $|\theta,\phi\rangle=\cos(\frac{\theta}{2})|0\rangle+e^{i\phi}\sin(\frac{\theta}{2})|1\rangle$. This state receives a unitary transformation $\ket{\theta,\phi}\rightarrow U_j(\theta_R,\delta)\ket{\theta,\phi}$ with $U_j(\theta_R,\delta)$ the operator for a $\theta_R$ rotation
about axis $j$  detuned by $\delta$. The fidelity of the transformed state with respect to the original state, averaged over the Bloch sphere, is   
\begin{equation}
F(\theta_R,\delta)=\frac{\int^{\pi}_{0}d\theta\, \sin(\theta)  \int^{2\pi}_{0} d\phi\,  |\langle\theta,\phi|U_j(\theta_R,\delta)\ket{\theta,\phi}|^2 }{4\pi}. \label{eq:crosstalk}
\end{equation}
The crosstalk error defined as $E_{\rm xt}=1-F(\theta_R,\delta)$  is
 shown in Fig. \ref{fig:state_fidelity} for several elements of $\mathcal C_1$. We see that the simple estimate of $\delta/\Omega\simeq \sqrt{15}$ for a $R_x(\pi)$ rotation is verified by the full calculation.  Since the error is minimized at different detunings for different rotations the detuning should be dynamically adjusted in tact with the gate being performed. In the demonstration described below we have simply used a fixed detuning
of $\delta/\Omega = \frac{2\pi\times 33~{\rm  kHz}}{2\pi\times 8.5~{\rm kHz}}=3.88\simeq\sqrt{15}$.

\begin{figure}
\includegraphics[width=0.9\columnwidth]{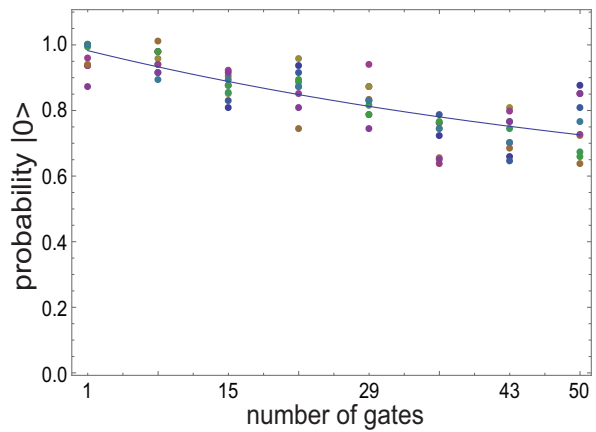}
\vspace{-.3cm}
\caption{(color online) RB data at Stark beam addressed site 31 for 10 RB sequences. Each data point is an average of 50 measurements.}
\label{fig:single_site}
\end{figure}

To characterize site selected gates  we applied 10 randomized Clifford
sequences as shown in Fig. \ref{fig:single_site}. Each sequence was truncated at 8 different lengths
\{1, 8, 15, 22, 29, 36, 43, 50\}.  Averaging over the  10
sequences yields  an average gate fidelity   $F^2=0.9923$, giving a per gate error which is about 4.5 times larger than for the array averaged global gates.
 We attribute this to fluctuations in the intensity and pointing stability of the Stark shifting beam resulting in deviations from the optimal detuning condition.

The crosstalk error at other sites was measured  by preparing them in  $\ket{1}$ and then measuring 
$P_{\ket{0}}=\frac{1}{2}-\frac{1}{2}(1-d_{\rm if})(1-d_{\rm xt})^{\ell}$ after each Clifford sequence was applied.
Dropping the Stark addressed site and sites whose loading
was poor
 yields an average $\left\langle d_{\rm xt}\right\rangle$. The array averaged background error on non-addressed sites 
 per Clifford was  $\left\langle E_{\rm xt}\right\rangle=\left\langle d_{\rm xt}\right\rangle/2=0.002(9)$. Due to the finite size of the Stark beam there was intensity overlap to nearest neighbor sites that was as high as 5\%, resulting in increased crosstalk compared to further away sites. The 
crosstalk values for the neighboring sites are given in Table \ref{tab:results}.   The average crosstalk error is comparable to the average error of global gates. However, this result was obtained for an initial state in the non targeted sites of $\ket{0}$ and therefore essentially corresponds to a spin flip error. It is to be expected that measurements with arbitrary initial states would yield higher errors (compare Figs. \ref{fig:state_fidelity}a) and b)). Ultimately, a slightly smaller Stark beam, and  larger values of $\delta/\Omega$ than have been demonstrated here should be used for effective crosstalk suppression.

In summary we have demonstrated high fidelity single qubit gate operations in a 2D array of neutral atom qubits. Using microwave pulses we perform either parallel gates on all qubits, or gates on single qubits selected by a Stark shifting beam. The results reported, together with the demonstration of two-qubit entanglement in the array using Rydberg blockade gates, which we will report on 
elsewhere\cite{Maller2015}, are a  step towards scalable quantum computing with neutral atom qubits.

This work was supported by the IARPA MQCO
program through ARO contract W911NF-10-1-0347.


%


\newpage

\widetext

\setcounter{page}{1}

\setcounter{table}{0}

\section{Supplementary material for Randomized benchmarking of single qubit gates in a 2D array of neutral atom qubits}

In this supplementary material we specify the microwave pulses used for each of the 24 gates in the single qubit Clifford group $\mathcal C_1$.
The group elements and our implementation are listed in Table \ref{tab.Clifford1}. Note that rotations of  $-\pi/2$ were implemented as $+3\pi/2$ rotations. We therefore used microwave pulses which were longer than necessary giving an average pulse area per Clifford gate of $\langle\theta\rangle_{\mathcal C_1}=7\pi/4.$ If all $3\pi/2$ operations had been replaced by $-\pi/2$ rotations this would have reduced the average pulse area to $55\pi/48$. As discussed in the main text the finite $T_2^\ast$ time accounts for about half of the observed gate error. This could be improved on simply by using the shorter $-\pi/2$ rotations instead of $3\pi/2$ rotations.

\begin{table}[!t]
\small
\begin{tabular}{|c|c|c|c|c||c|c|c|c|c|c|}
\hline
index  & $x$ axis & $y$ axis & $z$ axis& $U$& $R_3$  & $R_2$ & $R_1$&$\theta_{\rm total}$\\
\hline
$1$ & $I$ & $I$ & $I$& $\M{1}{0}{0}{1}$&-&-&-&0\\
$2$ & $I$ & $I$ & $\pi/2$& $e^{-\imath\pi/4}\M{1}{0}{0}{i}$&$R_x(\pi/2)$&  $R_y(\pi/2)$&  $  R_x(3\pi/2)$&$5\pi/2$\\ 
$3$ & $I$ & $I$ & $\pi$& $-i\M{1}{0}{0}{-1}$&- & $R_x(\pi)$& $ R_y(\pi)$&$2\pi$ \\ 
$4$ & $I$ & $I$ & $-\pi/2$& $e^{\imath\pi/4}\M{1}{0}{0}{-i}$ &$R_x(\pi/2)$&$ R_y(3\pi/2)$&$ R_x(3\pi/2)$&$7\pi/2$\\ 

$5$ & $I$& $\pi$ & $I$& $-1\M{0}{1}{-1}{0}$&-&-& $R_y(\pi)$&$\pi$\\ 
$6$ & $I$ & $\pi$ & $\pi/2$& $-e^{\imath \pi/4}\M{0}{1}{i}{0}$& $R_x(\pi/2)$&$ R_y(\pi/2)$ &$ R_x(\pi/2)$&$3\pi/2$\\ 
$7$ & $\pi$ & $I$ & $I$& $-i\M{0}{1}{1}{0}$ & - & - &$R_x(\pi)$&$\pi$\\ 
$8$ & $\pi$ & $I$ & $\pi/2$& $e^{-\imath\pi/4}\M{0}{1}{-i}{0}$ & $ R_x(\pi/2)$&$ R_y(3\pi/2)$&$ R_x(\pi/2)$&$5\pi/2$\\ 

$9$ & $\pi$ & $\pi/2$ & $I$& $\frac{-i}{\sqrt2}\M{1}{1}{1}{-1}$&- &$ R_x(\pi)$&$R_y(\pi/2)$&$3\pi/2$\\ 
$10$ & $I$ & $-\pi/2$ & $I$& $\frac{1}{\sqrt2}\M{1}{1}{-1}{1}$&-&-&$R_y(3\pi/2)$&$3\pi/2$\\ 
$11$ & $\pi/2$ & $I$ & $\pi/2$& $\frac{e^{-\imath\pi/4}}{\sqrt2}\M{1}{1}{-i}{i}$& - &$R_y(3\pi/2) $&$  R_x(\pi/2)$&$2\pi$\\ 
$12$ & $\pi/2$ & $\pi$ & $\pi/2$& $-\frac{e^{\imath\pi/4}}{\sqrt2}\M{1}{1}{i}{-i}$& - &$R_y(3\pi/2) $&$  R_x(3\pi/2)$&$3\pi$\\ 

$13$ & $\pi$ & $-\pi/2$ & $I$& $\frac{i}{\sqrt2}\M{1}{-1}{-1}{-1}$& - &$R_x(\pi) $&$  R_y(3\pi/2)$&$5\pi/2$\\ 
$14$ & $-\pi/2$ & $I$ & $\pi/2$& $\frac{e^{-\imath\pi/4}}{\sqrt2}\M{1}{-1}{i}{i}$& - &$R_y(\pi/2) $&$  R_x(3\pi/2)$&$2\pi$\\ 
$15$ & $I$ & $\pi/2$ & $I$& $\frac{1}{\sqrt2}\M{1}{-1}{1}{1}$& - &-&$  R_y(\pi/2)$&$\pi/2$\\ 
$16$ & $-\pi/2$ & $\pi$ & $\pi/2$& $\frac{e^{\imath\pi/4}}{\sqrt2}\M{1}{-1}{-i}{-i}$& - &$R_y(\pi/2) $&$  R_x(\pi/2)$&$\pi$\\ 

$17$ & $-\pi/2$ & $-\pi/2$ & $I$& $\frac{e^{-\imath\pi/4}}{\sqrt2}\M{1}{i}{-1}{i}$& - &$R_x(3\pi/2) $&$  R_y(3\pi/2)$&$3\pi$\\ 
$18$ & $-\pi/2$ & $\pi/2$ & $I$& $\frac{e^{\imath\pi/4}}{\sqrt2}\M{1}{i}{1}{-i}$& - &$R_x(3\pi/2) $&$  R_y(\pi/2)$&$2\pi$\\ 
$19$ & $-\pi/2$ & $\pi$ & $I$& $\frac{i}{\sqrt2}\M{1}{i}{-i}{-1}$& - &$R_x(3\pi/2) $&$  R_y(\pi)$&$5\pi/2$\\ 
$20$ & $-\pi/2$ & $I$ & $I$& $\frac{1}{\sqrt2}\M{1}{i}{i}{1}$& - &-&$  R_x(3\pi/2)$&$3\pi/2$\\ 

$21$ & $\pi/2$ & $-\pi/2$ & $I$& $\frac{e^{\imath\pi/4}}{\sqrt2}\M{1}{-i}{-1}{-i}$& - &$R_x(\pi/2) $&$  R_y(3\pi/2)$&$2\pi$\\ 
$22$ & $\pi/2$ & $I$ & $I$& $\frac{1}{\sqrt2}\M{1}{-i}{-i}{1}$& - &-&$  R_x(\pi/2)$&$\pi/2$\\ 
$23$ & $\pi/2$ & $\pi$ & $I$& $\frac{-i}{\sqrt2}\M{1}{-i}{i}{-1}$& - &$R_x(\pi/2) $&$  R_y(\pi)$&$3\pi/2$\\ 
$24$ & $\pi/2$ & $\pi/2$ & $I$& $\frac{e^{-\imath\pi/4}}{\sqrt2}\M{1}{-i}{1}{i}$& - &$R_x(\pi/2) $&$  R_y(\pi/2)$&$\pi$\\ 

\hline 
\end{tabular}
\caption{Elements of the Clifford group $\mathcal C_1$ for a single qubit. 
The notation $\theta$ in column 2,3 or 4 is  shorthand for $R_j(\theta)$ about the corresponding axis. The resulting operator $U$ is due to the sequence $U_x U_y U_z$, reading right to left.  
The operators have been factored so that the first nonzero element in the top row is unity. The microwave implementation $R_3 R_2 R_1$, reading right to left, used the rotations shown in the last three columns. These give the same Clifford operations as $U$, up to irrelevant global phase factors.  }
\label{tab.Clifford1}
\end{table}

\end{document}